\documentstyle[eqsecnum,preprint,aps,epsfig,floats]{revtex}
\newcommand{\Be}{\begin{equation}}
\newcommand{\Ee}{\end{equation}}
\newcommand{\be}{\begin{displaymath}}
\newcommand{\ee}{\end{displaymath}}
\newcommand{\Ba}{\begin{eqnarray}}
\newcommand{\Ea}{\end{eqnarray}}
\newcommand{\ba}{\begin{eqnarray*}}
\newcommand{\ea}{\end{eqnarray*}}
\newcommand{\Bml}{\begin{mathletters}}
\newcommand{\Eml}{\end{mathletters}}

\newcommand{\om}{\omega}
\newcommand{\BD}{B \rightarrow D}
\newcommand{\LL}{\Lambda_b \rightarrow \Lambda_c}

\begin{document}

\tighten

\thispagestyle{empty}

{\flushright{MZ-TH/99-12}
\flushright{IRB-TH-3/00}
\flushright{}
}

\vglue 2cm

\begin{center} \begin{Large} \begin{bf}
Exclusive-Inclusive Ratio of Semileptonic $\Lambda_b$-Decays
\end{bf} \end{Large} \end{center}
\vglue 0.35cm
{\begin{center}
\vglue 1.5cm
J.G.\ K\"{o}rner$^{\,1}$
\parbox{6.4in}{\leftskip=1.0pc
{\it Institut f\"{u}r Physik, Johannes
Gutenberg-Universit\"{a}t, D-55099 Mainz, Germany} }\\
\vglue 1.0cm
B.\ Meli\'{c}$^{\;2}$
\parbox{6.4in}{\leftskip=1.0pc
{\it Theoretical Physics Division, Rudjer Bo\v{s}kovi\'{c} Institute,
        HR-10002 Zagreb, Croatia} }
\end{center}}


\begin{center}
\vglue 3.0cm
\begin{bf} ABSTRACT \end{bf}
\end{center}
{
\noindent
We present theoretical evidence that the exclusive/inclusive ratio
of semileptonic $\Lambda_b$-decays exceeds that of semileptonic
B-decays where the experimental exclusive/inclusive ratio amounts to about 66\%.
We start from the observation that the spectator quark model provides
a lower bound on the leading order Isgur-Wise function of the $\LL$
transition in terms of the corresponding $ B \rightarrow D,D^* $ mesonic
Isgur-Wise function. Using experimental data for the $ B \rightarrow D,D^* $
mesonic Isgur-Wise functions this bound is established. Applying
a Bethe-Salpeter
model including spectator quark interactions and a QCD sum rule estimate of
the $\LL$ transition form factor which satisfy the spectator quark model
bound we predict the exclusive/inclusive ratio of semileptonic $\Lambda_b$ decay rates
to lie in a range between 0.81 and 0.92.
We also provide an
upper bound on the baryonic Isgur-Wise function which is determined from
the requirement that the exclusive rate should not exceed the inclusive
rate.
}

\renewcommand{\thefootnote}{1}
\footnotetext{e-mail: koerner@thep.physik.uni-mainz.de}
\renewcommand{\thefootnote}{\arabic{footnote}}
\addtocounter{footnote}{2}
\footnotetext{e-mail: melic@thphys.irb.hr}

\newpage

\section{Introduction}

In mesonic semileptonic $b \rightarrow c$ transitions, the exclusive
transitions to the ground state $S$-wave mesons $B \rightarrow D,D^*$ 
make up approximately $66 \%$ of the total 
semileptonic $B \rightarrow X_c$ rate \cite{PDG}. 
It  would then be  interesting to know what the corresponding semileptonic rate ratio
$\Gamma_{\Lambda_b \rightarrow \Lambda_c}/\Gamma_{\Lambda_b \rightarrow X_c}$
(termed $R_E$ in the following) is in semileptonic $\Lambda_b$-decays. This is an important
experimental issue since a knowledge of this ratio would greatly facilitate the
analysis of semileptonic $\Lambda_b$-decays. For example, if the
semileptonic $\Lambda_b$-decays were dominated by the quasi-elastic
exclusive channel $\Lambda_b \rightarrow \Lambda_c +l^- + 
\overline{\nu}_l$\hspace{4mm} this would
be of considerable help in the kinematical reconstruction
of their decays in as much as the $\Lambda_c$ baryon is easy to detect 
via its decay mode
$\Lambda_c \rightarrow p K^- \pi^+$.  Unfortunately nothing is known experimentally about this
ratio yet.

In this paper we attempt
to address the problem of a determining the exclusive/inclusive ratio $R_E$  
in semileptonic $\Lambda_b$-decays from a theoretical point of view by consulting
some model calculations which we critically scrutinize. We also attempt
to extrapolate from the experimentally known results in the meson sector
to the baryon sector.
 
As concerns the inclusive semileptonic rates of bottom mesons and bottom
baryons one is now reasonably confident that they
can be reliably calculated using the usual operator product 
expansion within HQET. 
The leading term in the OPE is given by the free heavy 
quark
decay rate which clearly is the same for baryons and mesons.
Radiative corrections to the free quark decay rate are quite large but
again are identical for mesons and baryons. Differences in the
inclusive semileptonic rates of mesons and the $\Lambda_b$ baryon 
set in only at ${\cal O}(1/m_b^2)$. They affect the mesonic and
$\Lambda_b$ rates differently since there is no  
chromomagnetic ${\cal O}(1/m_b^2)$ correction in the $\Lambda_b$ case. However,
since the chromomagnetic term contributes only at the  $3.7 \% $ level, 
the difference in the inclusive semileptonic rates for mesons and baryons 
is predicted to be quite small.

A much more difficult task is to get a reliable theoretical handle on the 
quasi-elastic exclusive semileptonic $\Lambda_b \rightarrow \Lambda_c$  rate.
There exist a number of theoretical calculations on the 
exclusive decay $\Lambda_b \rightarrow \Lambda_c +l^- + \overline{\nu}_l$ using
various model assumptions. They are of no great help since their
predicted rate values may differ by factors of up to three and it is not 
easy to judge the reliability of the various model assumptions 
that enter the calculation. Ideally one
would like to have model calculations that are valid both in the heavy meson and the
heavy baryon sector. If these model calculations give sensible results in 
the heavy meson sector, where they can be checked against data, one would
have more confidence in their predictions for the heavy baryon sector.

The paper is structured as follows. In Sec.II we take a first look at
the leading order rate formula for the exclusive semileptonic decays of
B mesons and $\Lambda_b$ baryons in order to get a semiquantitative
handle on the relative size of their rates. The analysis is refined
in Sec.III for $\Lambda_b$ baryons where $1/m_b$ and $1/m_b^2$
corrections and renormalization effects are included. In Sec.IV we
recapitulate the calculation of the semileptonic inclusive decay
rates. The results of the Sections III and IV are brought together in Sec.V
where we discuss the exclusive/inclusive ratio $R_E$ of semileptonic
$\Lambda_b$ decays. We present numerical results on the exclusive/inclusive
ratio for various models and give our best estimate of this ratio. In
Sec.VI we classify the possible non-exclusive final
states that will have to fill the gap between the exclusive and inclusive rates
in semileptonic $\Lambda_b$ decays.
Sec.VII, finally, contains our conclusions.    

\section{The heavy quark limit}

For a quick first appraisal of the question of how the exclusive
semileptonic decays of mesons and baryons are related we turn to
the heavy quark limit and list 
the leading order semileptonic rate formulae for the
$B \rightarrow (D + D^*)$ and $\Lambda_b \rightarrow \Lambda_c$ transitions. 
One has \cite{PDG,KKP}
\begin{equation}
\frac{d\Gamma \left \{ \rm  meson \atop baryon \right \} }{d \omega} = 
\frac{G_F^2 |V_{bc}|^2 M_1^5}{12 \pi^3} r^3 
\sqrt{\omega^2 -1} \left (3\omega (1+r^2) - 2 r (2\omega^2+1) \right )
\left \{ \frac{\omega + 1}{2} |F_{\rm meson}(\omega)|^2 \atop
|F_{\rm baryon}(\omega)|^2  \right \}  , 
\label{gen}
\end{equation}
where $r = M_2/M_1$ and $\om = (M_1^2 + M_2^2 -q^2)/2 M_1 M_2$. 
$F_{\rm meson}(\om )$ and $F_{\rm baryon}(\omega)$ are 
the leading order Isgur-Wise transition form factors for the $\BD$,$D^*$ and
$\LL$ transitions, respectively. 
Throughout the paper we refer to $M_1$ and $M_2$ as the masses of the
initial and final particles in the semileptonic decay process. 

The free
heavy quark decay rate (or leading order parton model rate)
which we need later on is simply obtained
by replacing the particle masses in (\ref{gen}) by the corresponding quark masses
and setting the curly bracket in (\ref{gen}) to one,
i.e. by taking the current coupling in the $\Lambda_b \rightarrow \Lambda_c$ case
to be point-like.
We shall encounter
the integrated parton model rate for the $\Lambda_b \rightarrow \Lambda_c$ 
case again in Sec.IV. 
Finally, in the heavy quark limit
one has to  determine the final meson mass $M_2$ by taking the weighted average 
$ \overline{M}_D  = 1/4 (M_D + 3 M_{D^*})= 1.973 \,\rm GeV$ with
$M_D= 1.869\,\rm GeV$ and $M_{D^*}=2.010\,\rm GeV$. 
For the pseudoscalar
bottom mass we take $ M_B = 5.279 \,\rm GeV$.
For the $\Lambda_Q$-baryon masses
we use $M_{\Lambda_b}=5.624 \,\rm GeV$ and $M_{\Lambda_c}=2.285 \,\rm GeV$.

When trying to compare the two rates in (\ref{gen}) one identifies two
main determining factors which counteract each other. On the one hand
one has the form factor expressions in the curly bracket which tend to enhance
the mesonic rate due to the multiplicative factor $(\omega+1)/2$ in 
the mesonic case.
Also, according to 
common prejudice the baryon form
factor falls off more rapidly than the mesonic form factor. On the other
hand one has the overall mass factor $M_1^5 r^3$ which enhances the
baryonic rate because $M_1^5 r^3 = 214.16\, \rm GeV^5$  and
$M_1^5 r^3 = 377.37 \, \rm GeV^5$ in the mesonic and baryonic cases, respectively.

It is evident that the choice of mesonic and baryonic Isgur-Wise functions plays
a crucial role when comparing the two rates. As has been emphasized
before, there exist some experimental knowledge on the mesonic
Isgur-Wise function but nothing is known experimentally about the baryonic
Isgur-Wise function yet
\footnote{The only available
experimental result is from a preprint version of a DELPHI-analysis
\cite{DELPHI}. This paper
quotes a value of $\rho^2=1.81^{\displaystyle +0.70}_{\displaystyle -0.67} \pm 0.32$ for the slope
of the baryonic Isgur-Wise function. However, since this paper has never been
published, we shall not use their result in our analysis.}.

For quick reference it is sometimes convenient to characterize the fall-off
behaviour of the Isgur-Wise functions by expanding it around
the zero recoil point where one has the zero recoil normalization condition
$F(1)= 1$. Keeping terms up to second order in this expansion one has
\Be
F(\om)= F(1)[ 1 - \rho^2 (\om -1) + c\, (\om -1)^2 + ... \,],
\label{exp}
\Ee
where  the coefficients $\rho^2$ and $c$ are called the slope parameter and
the convexity parameter, respectively. The slope
parameter is frequently used to characterize the fall-off behaviour of the 
Isgur-Wise function. 
The expansion (\ref{exp}) is useful if one
studies the physics close to zero threshold but may give misleading results
when calculating rates because
the spectral weight function multiplying the form factor functions
is essentially determined by the square root factor
$\sqrt{\om^2-1}$ in (\ref{gen})
and  is therefore  strongly
weighted towards the end of the spectrum. It goes
without saying that the slope and convexity parameters are in general 
different for 
the mesonic and baryonic Isgur-Wise functions. 

\begin{figure}[ht]  
\centerline{\epsfig{file=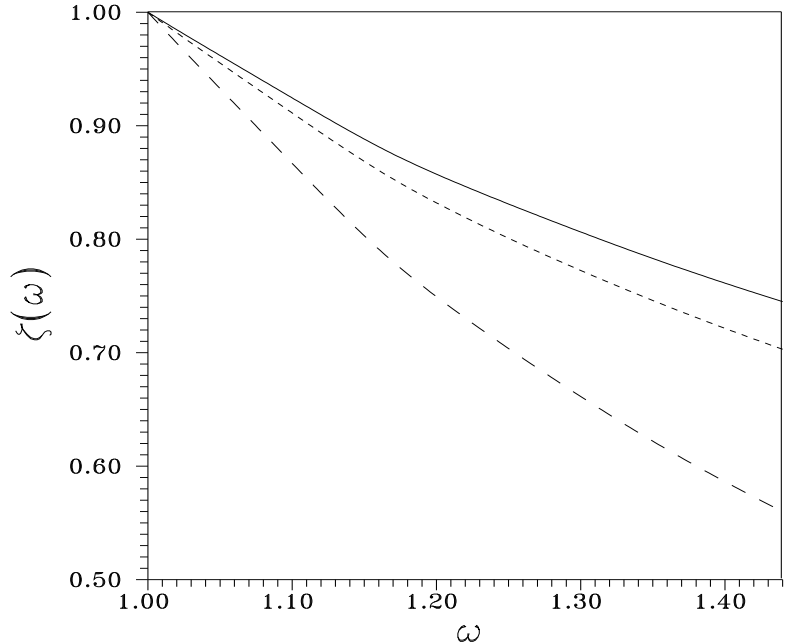,height=9cm,width=10.5cm,silent=}}
\caption{Leading order baryon form factor $F_{\rm baryon}
(\om)$ in the dynamic Bethe-Salpeter
model of \protect\cite{IKLR} (denoted by $\zeta(\om)$ in this figure). 
a) Noninteracting light quarks (long-dashed line) and 
b) interacting light quarks with the range parameter 
$\Lambda_B = 500\, {\rm MeV}$ (solid line) and 
$\Lambda_B = 355\, {\rm MeV}$ (short-dashed line).}
\label{f:fig1}
\end{figure}

In order to proceed with our first appraisal of the magnitude of the exclusive
mesonic and baryonic semileptonic rates we appeal to the spectator
quark model where the mesonic and baryonic Isgur-Wise functions become
related to one another \cite{KKP,IKLR}. In the spectator quark model one has
\Be
F_{\rm baryon}(\omega) = \frac{\omega + 1}{2} |F_{\rm meson}(\omega)|^2 .
\label{spec}
\Ee
Explicit calculations show that the baryonic form factor is considerably
underestimated by the spectator relation (\ref{spec}). Nevertheless, 
the spectator
relation (\ref{spec}) may still serve as an
effective lower bound on the baryonic form factor.
 
The physical picture behind the spectator quark model relation is quite simple. In the heavy baryon
case there are two light spectator quarks that need to be accelerated
in the current transition compared to the one spectator quark
in the heavy meson transition. Thus the baryonic form factor is determined
in terms of the square
of the mesonic form factor. The factor $(\frac{\omega + 1}{2})$ is a relativistic
factor which insures the correct threshold behaviour of the baryonic 
form factor in the crossed
$e^+ e^-$-channel \cite{KKP,IKLR}.
   
In \cite{IKLR} the relation between heavy meson
and heavy baryon form factors was investigated in the context of
a dynamical Bethe-Salpeter (BS) model. The above spectator quark model
relation (\ref{spec}) in fact emerges when the
interaction between the light quarks in the heavy baryon is switched
off in the BS-interaction kernel. In the more realistic situation when
the light quarks interact with each other,  
the heavy baryon form factor becomes flatter, i.e.
the spectator quark model form factor may be used to bound 
the heavy baryon form factor from below. In Fig.1 we reproduce from 
\cite{IKLR} the
$\omega$ dependence of the spectator quark model form factor and that of
two representative form factors with the interaction between the
light quarks included. The starting point in \cite{IKLR} is a mesonic
form factor
with a slope of $\rho^2_{\rm meson} = 1 $ which, according to (\ref{spec}),
leads to a spectator  
form factor with a slope of $\rho^2_{\rm baryon} = 1.5 $. The spectator
form factor is the lowest form factor shown in Fig.1.  
The interaction between the light quarks was introduced through a harmonic
oscillator type kernel in the BS-equation. The two upper form factor curves
in  Fig.1 correspond to two different choices of the oscillator strength with
which the light quarks interact or,
equivalently, correspond to two different choices of the size parameter in the 
oscillator wave function. The interaction type form factors
in Fig.1 have slopes of $\rho^2_{\rm baryon} = 0.81$ (solid line) and 0.97 (short-dashed line) 
\cite{IKLR}. They
are considerably flatter than the spectator quark model form factor.

We shall now calculate exclusive rates for mesonic and
baryonic transitions according to (\ref{gen}) using the spectator model
relation (\ref{spec}). According to what was said before, the baryonic rate
calculated in this way must subsequently
be adjusted upward according to the analysis of \cite{IKLR}. We went
to considerable lengths in explaining the
results of \cite{IKLR} because we want to emphasize that the outcome
of the baryonic rate estimate using the spectator quark model relation (\ref{spec})
must be viewed as providing only lower bounds on
the true quasi-elastic baryonic rate. For the mesonic form factor we use the
world average of the slope 
$\rho_{\rm meson}^2$ obtained
by combining results from $B \rightarrow D^*$ and 
$B \rightarrow D$ transitions, $\rho_{\rm meson}^2 = 0.70$ \cite{Drell}
\footnote{The CLEO Coll.
also attempted a linear plus quadratic 
fit to the data, but the data was not good enough to determine the
convexity parameter $c$ of the meson form factor with any accuracy.}.
Using $V_{cb}=0.038$, a linear meson form factor with the above slope, 
a baryonic form
factor according to the spectator relation (\ref{spec}) and the rate
formulae (\ref{gen}) 
one obtains
$\Gamma_{\rm meson}= 5.30 \cdot 10^{10} s^{-1}$ and 
$\Gamma_{\rm baryon}= 5.04 \cdot 10^{10} s^{-1}$. 
As has been emphasized before the baryonic rate 
$\Gamma_{\rm baryon}= 5.04 \cdot 10^{10} s^{-1}$ has to be adjusted
upward in the more realistic situation of interacting light quarks. Looking at the
model calculation \cite{IKLR} for guidance, the increment in rate 
going from
noninteracting (spectator) to interacting quarks is 1.28 and 1.37, respectively,
for the two choices of oscillator strengths analyzed in \cite{IKLR}. Adjusting the
above baryonic rate  accordingly our leading order
estimate of the baryonic rate is thus
$\Gamma_{\rm baryon}= (6.45 \div 6.90) \cdot 10^{10} s^{-1}$.
Starting from a mesonic exclusive/inclusive ratio of $66\%$ and assuming
equal  inclusive semileptonic rates for bottom baryons and mesons, which is
sufficiently accurate for our semiquantitative calculation, our estimate
for the  exclusive/inclusive ratio in semileptonic $\Lambda_b$ decays is
$R_{E} = (80-86)\% $. This is considerably larger than the mesonic
exclusive/inclusive ratio $R_E \approx 66\%$.  

Up to this point our semiquantitive analysis was done to leading order in HQET.
How would finite mass effects affect our previous conclusions? One way of
improving the previous analysis in the meson sector is to  insert physical masses
in the rate expression
(\ref{gen}), thereby 
including part of the $1/m_Q$-corrections to (\ref{gen}). To do this we
need to disentangle the 
$\BD$ and $\BD^*$ rates in (\ref{gen}). One has \cite{PDG}:
\begin{equation}
\frac{d\Gamma \left (\BD  \right ) }{d \omega} = \frac{G_F^2 |V_{bc}|^2 M_1^5}{48 \pi^3} r^3 (1+r)^2
(\omega^2 -1)^{3/2} |F_{\rm meson}(\omega)|^2 
\nonumber
\end{equation}
and 
\begin{equation}
\frac{d\Gamma \left (\BD^*  \right ) }{d \omega} = \frac{G_F^2 |V_{bc}|^2 M_1^5}{48 \pi^3} r^3 
\sqrt{\om^2-1} (\omega +1) \left [ (1-r)^2(\om+1) + 4 \om (1-2 \om r + r^2) 
\right ]
|F_{\rm meson}(\omega)|^2 .
\nonumber
\end{equation}

Using now the world average of $(\rho^{2}_{\rm meson})^{\BD} = 0.66$  
and $(\rho^{2}_{\rm meson})^{\BD^*} = 0.71$ with  
linear form factors and taking 
physical $D$ and $D^*$ masses 
one finds
$\Gamma_{B \rightarrow D}= 1.39 \cdot 10^{10} {\rm s}^{-1}$ and 
$\Gamma_{B \rightarrow D^*}= 3.90 \cdot 10^{10} {\rm s}^{-1}$ giving a total
mesonic rate of 
$\Gamma_{B \rightarrow D + D^*}= 5.30 \cdot 10^{10} {\rm s}^{-1}$. 
It fully agrees with the above result
and therefore leaves the aforegoing conclusions intact.

Continuing with our discussion on the contributions of nonleading effects
in the $1/m_Q$-expansion we now turn to results of some model calculations
in order to find out how nonleading effects may affect the above conclusions.
In the mesonic sector Neubert and Rieckert \cite{neubert}
analyzed an infinite
momentum frame model and
found that ${\cal O}(1/m_Q)$ effects raise
the $B \rightarrow D$ and $B \rightarrow D^*$ rates by $15.7\%$ and $0.5\%$,
respectively, resulting in a rise of $4.4\%$ for the total $D + D^*$ rate. 
Using a
similar infinite momentum frame model K\"onig et al. find
that the ${\cal O}(1/m_Q)$ effects raise the semileptonic 
$\Lambda_b \rightarrow \Lambda_c$ rate
by $3\%$ \cite{KKKK} which is quite close to the $4.4\%$ found in \cite{neubert} 
in the
bottom meson case. Judging from these model calculations our 
leading order comparison
of the mesonic and baryonic rates and the conclusions drawn from it do not
seem to be much affected by ${\cal O}(1/m_Q)$ corrections.
  
There also exist estimates of ${\cal O}(1/m_Q^2)$ corrections in the literature.
Faustov and Galkin et al. use a relativistic quark model based on 
the quasipotential
approach \cite{rudolf}. They quote exclusive/inclusive branching ratios of 
$(13.5 + 3.3 - 1.4)\%$
and $(39.1 + 6.5 -3.9)\%$ for semileptonic 
$B \rightarrow D$ and $B \rightarrow D^*$ rates,
where the  second and third numbers refer to the 
${\cal O}(1/m_Q)$ and ${\cal O}(1/m_Q^2)$
corrections, respectively. The ${\cal O}(1/m_Q)$ corrections in this model
are considerably
larger than in the infinite momentum frame models. 
Ivanov et al. investigated the role of
finite mass effects in semileptonic $\Lambda_b \rightarrow \Lambda_c$ decays 
without
taking recourse to the heavy mass expansion. 
They found an overall rate reduction of
$9\%$ relative to the infinite mass result \cite{finite}. 
In the analysis of the present paper
presented in Sec.III
we obtain $\approx +5\%$ and $\approx -7\%$ for the ${\cal O}(1/m_Q)$ and
${\cal O}(1/m_Q^2)$ corrections to the $\Lambda_b \rightarrow \Lambda_c$ decays,
respectively. From all these model calculations one learns that 
the ${\cal O}(1/m_Q)$
corrections tend to increase the rates whereas the  
${\cal O}(1/m_Q^2)$ corrections
tend to decrease the rates, for both heavy meson and heavy baryon decays.         
Again, our leading order estimate of the relative size of the 
exclusive/inclusive ratios of bottom mesons and
baryons is not likely to be affected much by including  
also ${\cal O}(1/m_Q^2)$ effects.
The same holds true for renormalization effects of the weak current  which 
affect the
bottom baryon and bottom meson amplitudes equally and 
therefore drop out in the ratio
of exclusive semileptonic bottom meson and baryon decays.

The conclusion drawn in this section on the predominance of the  
exclusive/inclusive ratio of semileptonic $\Lambda_b$-decays
over that of semileptonic B-decays
carries over to 
the more sophisticated analysis 
of the next sections where we include $1/m_Q$ and $1/m_Q^2$ effects, 
and radiative corrections.




\section{Exclusive Semileptonic Rate  
$\lowercase{\Lambda_b \rightarrow 
\Lambda_c + l^- + \overline{\nu}_l}$ }

It is most convenient to represent the differential decay rate in terms of 
the helicity 
amplitudes of the process. One has (we take leptons to be massless)
\cite{KKP,KK}
\Be
\frac{d\Gamma \left (\LL  \right ) }{d \omega} = 
\frac{G_F^2 |V_{bc}|^2 }{96 \pi^3} 
\frac{q^2 M_2^2 \sqrt{\om^2-1}}{M_1} \left ( |H_{\frac{1}{2}1}|^2 + 
|H_{-\frac{1}{2}-1}|^2 + |H_{\frac{1}{2}0}|^2 + |H_{-\frac{1}{2}0}|^2 \right ).
\label{gen3}
\Ee

The helicity amplitudes are in turn related to the invariant amplitudes of 
the process via
\Ba
\sqrt{q^2} H_{\frac{1}{2}0}^{V,A} &=& \sqrt{2 M_1 M_2 (\om \mp 1)} \left (
(M_1 \pm M_2) f_1^{V,A} \pm M_2 (\om \pm 1)  f_2^{V,A}  \pm M_1 (\om \pm 1)  
f_3^{V,A}
\right ) , \nonumber \\
H_{\frac{1}{2}1}^{V,A} &=& -2 \sqrt{M_1 M_2 (\om \mp 1)} f_1^{V,A} , 
\label{hel}
\Ea
where the invariant amplitudes are defined by 
\Ba
< \Lambda_c (v_2) | J_{\mu}^V |  \Lambda_b (v_1) > &=& 
\overline{u}_c(v_2)( f_1^V \gamma_{\mu} + f_2^V v_{1 \mu} + f_3^V v_{2 \mu} )u_b(v_1),  \nonumber \\
< \Lambda_c (v_2) | J_{\mu}^A |  \Lambda_b (v_1) > &=& 
\overline{u}_c(v_2)(f_1^A \gamma_{\mu} + f_2^A v_{1 \mu} + f_3^A v_{2 \mu} )\gamma_5 u_b(v_1) .
\Ea

The total helicity amplitudes finally are given by 
\Be
H_{\lambda_2\lambda_W} = H_{\lambda_2\lambda_W}^V - H_{\lambda_2\lambda_W}^A ,
\Ee
where the choice of of the relative minus sign between the vector and axial
vector helicity amplitudes reflects the $(V-A)$ structure of the
$b \rightarrow c$ current transition.
The $H_{\lambda_2\lambda_W}^{V,A}$ are the helicity amplitudes for the vector 
($V$) and axial vector ($A$) current 
induced transition in the decay $1/2^+ \rightarrow 1/2^+ + 
W_{off-shell}^-$ with $\lambda_2$ and $\lambda_W$ being the 
helicities of the final state baryon and the $W$ boson, respectively.

The remaining helicity amplitudes are related to the above helicity amplitudes 
(\ref{hel}) by parity.
One has
\Be
H_{-\lambda_2-\lambda_W}^{V,A} = \pm H_{\lambda_2\lambda_W}^{V,A} .
\Ee

It is well known that the complexity of the  form factor structure  exemplified
by the set of six form factors $f_i^{V,A} \,(i=1,2,3)$ is considerably 
reduced in HQET. Working up to 
${\cal O}(1/m_Q)$ in HQET and including also ${\cal O}(\alpha_s)$ corrections one finds \cite{KKP}
\Ba
f_1^V(\om) &=& F(\om) +  \left ( \frac{1}{2 M_1} + 
\frac{1}{2 M_2} \right ) \left (\eta (\om) + \overline{\Lambda} F(\om) \right ) 
+ \frac{\alpha_s(\overline{m})}{\pi} v_1(\omega,\lambda)  F(\om),  \nonumber \\
f_2^V(\om) &=&  F(\om)\left (-\frac{1}{M_2} \frac{1}{\om + 1} 
\overline{\Lambda} 
- \frac{\alpha_s(\overline{m})}{\pi} v_2(\omega) \right ), \nonumber \\
f_3^V(\om) &=&  F(\om)\left (-\frac{1}{M_1} \frac{1}{\om + 1} \overline{\Lambda} 
- \frac{\alpha_s(\overline{m})}{\pi} v_3(\omega) \right ), \nonumber \\
f_1^A(\om) &=& F(\om) + \left ( \frac{1}{2 M_1} + 
\frac{1}{2 M_2} \right ) \left ( \eta (\om) + \overline{\Lambda} F(\om)
\frac{\om -1}{\om + 1} \right ) 
+ \frac{\alpha_s(\overline{m})}{\pi} a_1(\omega,\lambda)  F(\om), \nonumber \\
f_2^A(\om) &=&  F(\om)\left (-\frac{1}{M_2} \frac{1}{\om + 1} 
\overline{\Lambda} 
- \frac{\alpha_s(\overline{m})}{\pi} a_2(\omega) \right ), \nonumber \\
f_3^A(\om) &=&  F(\om)\left (\frac{1}{M_1} \frac{1}{\om + 1} \overline{\Lambda} 
- \frac{\alpha_s(\overline{m})}{\pi} a_3(\omega) \right ).
\label{ff}
\Ea

The ${\cal O}(\alpha_s)$ corrections to the form factors have been 
taken from \cite{Neubert}.
They result from the ${\cal O}(\alpha_s)$ vertex correction to the current-induced 
$b \rightarrow c$ transition \cite{Pasch}. The infrared singularity 
is regularized by the introduction
of a fictitious gluon mass which is taken to be 
$\lambda = 0.2 \; {\rm GeV}$. At zero recoil,
where the vertex correction is infrared finite, 
the renormalization is independent of the
gluon mass regulator. However, away from zero recoil,  
the $\alpha_s$-correction functions
$v_1(\om, \lambda)$ and $a_1(\om, \lambda)$  depend on the gluon mass regulator.
This introduces a certain amount of model dependence in the renormalization procedure.
The above value of the gluon mass was chosen according to the expectation that the
exchange of virtual gluons in the vertex correction should be cut-off at frequencies
$k^0 \sim 1/R$ with $R \sim 1$ fm being a typical hadronic scale.  
The argument of the $\alpha_s$ 
coupling, $\overline{m}$, is taken such that effects of 
higher order terms $(\alpha_s \ln(m_b/m_c))^n$ 
are minimized, $\overline{m} = 2 m_b m_c/(m_b + m_c) \simeq 2.31\, {\rm GeV}$. 
In this way one 
avoids the use of the renormalization group improved summation of 
leading logarithms, which 
has been proven as inconsistent \cite{Neubert}.
 
The binding energy of the $\Lambda_b$ is denoted by $\overline{\Lambda}$
which we take to be 0.6 GeV. 
The form factor function $\eta(x)$ results from the nonlocal contribution of the
kinetic energy term of the
$1/m_Q$ corrected HQET Lagrangian. It  has been calculated in 
two different model approaches 
and has found to be negligibly small \cite{KKKK,DHHL}. Therefore, 
we can safely drop its contribution in the following. 
By neglecting the form factor $\eta(x)$ in the 
${\cal O}(1/m_Q)$ result (\ref{ff}) the differential
rate (\ref{gen3}) is proportional to the square of the leading order
Isgur-Wise function $F(\omega)$ with its zero recoil normalization $F(1)=1$. In this
way we can meaningfully compare our results with the leading order results
of other model calculations as will be done in Sec.V.

It is well known that ${\cal O}(1/m_Q^2)$ corrections to the unit zero recoil
normalization of Isgur-Wise functions can be substantial. For example,
by evaluating zero recoil sum rules, the authors of Refs.\cite{SUV} obtain
\Ba
F(1)_{B \rightarrow D} = 0.98 \pm 0.07 ,  \nonumber \\
F(1)_{B \rightarrow D^*} = 0.91 \pm 0.03 . 
\Ea
in the mesonic case. In the $\Lambda_b \rightarrow \Lambda_c$
case the zero recoil sum rule
gives a bound on the zero recoil value of the sole remaining form factor function $f_1^A$.
The (unrenormalized) zero recoil sum rule
leads to $f_1^A(1) \le (1- 0.165 \,\mu_{\pi}^2/{\rm GeV^2})^{1/2}$ \cite{KP}.
Using $\mu_{\pi}^2= 0.5\,{\rm GeV^2}$ [18,19], $f_1^A(1)$ must be smaller than
0.958. For definiteness we take a value close to the upper
bound
\Be
f^A_1(1) = 0.95 .
\Ee
This value is nicely corroborated by the 
finite heavy quark mass calculation of \cite{finite}
where one finds $f^A_1(1) = 0.97$.

Nothing is known about the size of the ${\cal O}(1/m_Q^2)$ corrections
to $\Lambda_b \rightarrow \Lambda_c$ away from zero recoil, 
except that they can be parametrized in terms of ten new $\om$-dependent 
form factors and one 
new dimensionful constant \cite{falk1}, the magnitude and functional forms of 
which are not known. The lack of knowledge about the ${\cal O}(1/m_Q^2)$ corrections
away from zero recoil prevents us from their exact treatment. On the other hand
the size of the ${\cal O}(1/m_Q^2)$ correction at zero recoil is a clear
indication that the ${\cal O}(1/m_Q^2)$ corrections cannot be neglected. We shall
therefore adopt the following strategy. We smoothly extrapolate from the
${\cal O}(1/m_Q^2)$ information at zero recoil to the whole $\omega$-range.
The appropiate amplitude for this extrapolation is the axial 
vector current $S$-wave amplitude
the zero recoil value of which is determined by the zero recoil sum rules. We thus
multiply the axial vector current $S$-wave amplitude everywhere by its zero recoil value
$f^A_1(1) = 0.95$. It is clear that the ${\cal O}(1/m_Q^2)$ corrections at zero recoil are
exactly included in this approach. The lack of knowledge
about the ${\cal O}(1/m_Q^2)$ corrections to the other partial wave amplitudes
leaves us no choice but to leave them untreated. With this in mind  
it is gratifying to note that the
$S$-wave contribution dominates the quasielastic rate. For example, using the standard form
factor (\ref{QCD}) with $\rho_B^2=0.75$ in a leading
order calculation one finds that the $S$-wave contribution amounts to $\simeq$ 66\%
of the total semileptonic rate.

In order to set up our procedure of how to incorporate the 
${\cal O}(1/m_Q^2)$ corrections we define the relevant vector current ($V$)
and axial vector current ($A$) partial wave amplitudes 
$A^{V,A}_{LS}$ in terms of the helicity amplitudes
$H_{\lambda_2\lambda_W}^{V,A}$. $L$ denotes the orbital
angular momentum of the final state and ${S}={J}_{{\rm current}}+{S}_{\Lambda_c}$
is the sum of the final state spin angular momenta where $J_{{\rm current}}=1$
in the zero lepton mass case that we are considering here.  One has

\Ba
A^V_{1\frac{1}{2}} &=& - \sqrt{\frac{2}{3}} H^V_{\frac{1}{2}0}
- \sqrt{\frac{4}{3}} H^V_{\frac{1}{2}1} \nonumber \; , \\
A^V_{2\frac{3}{2}} &=& - \sqrt{\frac{4}{3}} H^V_{\frac{1}{2}0}
+ \sqrt{\frac{2}{3}} H^V_{\frac{1}{2}1}\nonumber \; , \\
A^A_{0\frac{1}{2}} &=& \sqrt{\frac{2}{3}} H^A_{\frac{1}{2}0}
- \sqrt{\frac{4}{3}} H^A_{\frac{1}{2}1} \nonumber \; , \\
A^A_{2\frac{3}{2}} &=& \sqrt{\frac{4}{3}} H^A_{\frac{1}{2}0}
+ \sqrt{\frac{2}{3}} H^A_{\frac{1}{2}1}\, .
\Ea
By substituting invariant form factors according to (\ref{hel}) one can
verify the correct threshold behaviour of the partial wave amplitudes, {\it  i.e.}
$A^V_{1\frac{1}{2}},A^V_{1\frac{3}{2}} \sim (\omega - 1)^{1/2}$,
$A^A_{0\frac{1}{2}}\sim (\omega - 1)^0$ and
$A^A_{2\frac{3}{2}} \sim (\omega - 1)^1$.

According to the above strategy we now incorporate the ${\cal O}(1/m_Q^2)$
corrections by multiplying the $S$-wave amplitude $A^A_{0\frac{1}{2}}$ 
by the ${\cal O}(1/m_Q^2)$ zero recoil correction
$f^A_1(1)=0.95$. Thus we write
\Be
A^A_{0\frac{1}{2}}
\rightarrow f_1^A(1) A^A_{0\frac{1}{2}}= A^A_{0\frac{1}{2}} +
( f_1^A(1) -1)A^A_{0\frac{1}{2}}
\label{zero}
\Ee
For the first term on the $r.h.s.$ of (\ref{zero}) we substitute the  ${\cal O}(1/m_Q)$
result according to (\ref{ff}). Contrary to this we use only the leading order result for
$A^A_{0\frac{1}{2}}$ in the second term of (\ref{zero}) since it is already being multiplied by the 
${\cal O}(1/m_Q^2)$ factor $( f_1^A(1) -1)$. Including also the $A^A_{2\frac{3}{2}}$
partial wave amplitude the 
leading order expressions for the axial vector partial wave amplitudes read
\Ba
A^A_{0\frac{1}{2}} &=& \frac{1}{\sqrt{q^2}} \frac{2}{\sqrt{3}}
\sqrt{M_1M_2(\omega+1)}F(\omega) \left [ (M_1 - M_2 + 2\sqrt{q^2})(1 + 
\frac{\alpha_s(\overline{m})}{\pi} a_1(\om,\lambda)) \right . \nonumber \\
& & \left . - \frac{\alpha_s(\overline{m})}{\pi}(\om -1)
(M_2 a_2(\om) + M_1 a_3(\om)) \right ] \nonumber \\
A^A_{2\frac{3}{2}} &=& \frac{1}{\sqrt{q^2}} \frac{2\sqrt{2}}{\sqrt{3}}
\sqrt{M_1M_2(\omega+1)} F(\omega) \left [ (M_1 - M_2 - \sqrt{q^2})(1+
\frac{\alpha_s(\overline{m})}{\pi} a_1(\om,\lambda)) \right . \nonumber \\
& & \left .
 - \frac{\alpha_s(\overline{m})}{\pi}(\om -1)
(M_2 a_2(\om) + M_1 a_3(\om)) \right ]
\label{ls}
\Ea

Putting everything together we arrive at the differential rate. One obtains
\Ba
\frac{d\Gamma \left (\LL  \right ) }{d \omega} &=& \frac{G_F^2 |V_{bc}|^2 }
{48 \pi^3} 
\frac{q^2 M_2^2 \sqrt{\om^2-1}}{M_1}
\left \{ |H^V_{\frac{1}{2}1}|^2 + 
|H^V_{\frac{1}{2}0}|^2 + |H^A_{\frac{1}{2}1}|^2 + |H^A_{\frac{1}{2}0}|^2
\right . \nonumber \\ 
& &  \left . \hspace{-2cm}+  \left( (f^A_1(1))^2 -1 \right)\frac{2}{3}
\frac{M_1 M_2}{q^2} (\om + 1) F^2( \om ) ( M_1 - M_2 + 2 \sqrt{q^2} ) 
\right . \nonumber \\
& & \left . \hspace{-2cm} \left [ 
( M_1 - M_2 + 2 \sqrt{q^2} ) (1 + 2 \frac{\alpha_s(\overline{m})}{\pi}
 a_1(\om, \lambda))
- 2 \frac{\alpha_s(\overline{m})}{\pi}(\omega -1)
(M_2 a_2(\om) + M_1 a_3(\om)) \right ]
\right \}.
\label{gen4}
\Ea
In Eq.(\ref{gen4}) the first line incorporates the ${\cal O}(1)$ and
${\cal O}(1/m_Q)$ contributions including the radiative corrections as specified
by (\ref{hel}) and by (\ref{ff}).
The second and third lines comprise the 
${\cal O}(1/m_Q^2)$ corrections (including radiative corrections) as described before.
Since our aim is to compare our exclusive rate with the inclusive ${\cal O}(\alpha_s)$
rate written down in Sec.IV we have only retained radiative corrections up to 
${\cal O}(\alpha_s)$ in the exclusive rate (\ref{gen4}) for consistency reasons.

\begin{figure}
 \centerline{\epsfig{file=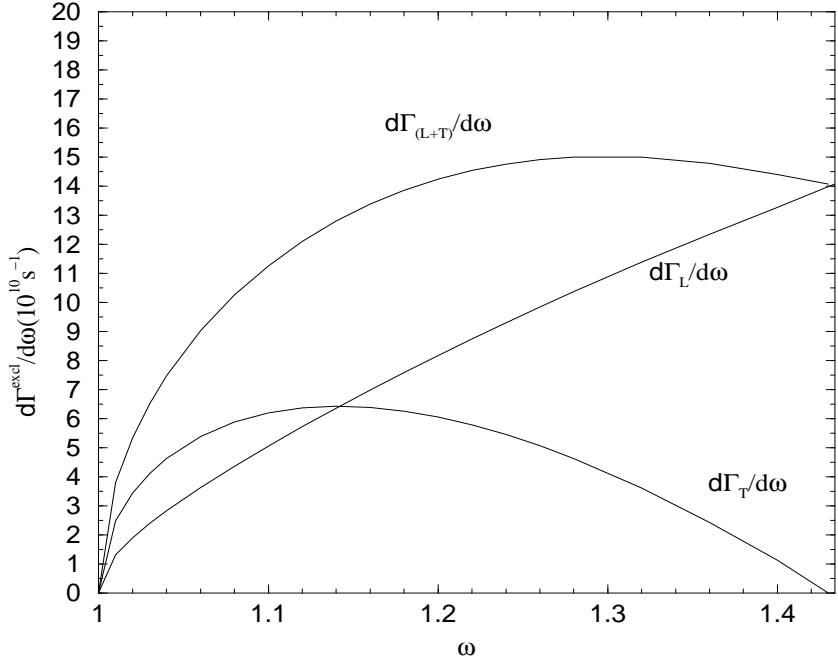,height=9cm,width=11cm,silent=}}
 \caption{$\om$ spectrum of the exclusive decay rate and partial 
rates of longitudinal and transversal
transitions calculated with the standard form factor 
(\protect{\ref{QCD}}) using $\rho_B^2=0.75$.} 
\label{f:fig2}
\end{figure}

For the sake of completeness we separately list the ${\cal O}(1/m_Q^2)$
zero recoil corrections for the longitudinal and transverse pieces
of the axial vector contributions. They are needed for the transverse/longitudinal
separation shown in Fig.2. One has
\Ba
|H^A_{\frac{1}{2}0}|^2 \rightarrow |H^A_{\frac{1}{2}0}|^2 +
((f^A_1(1))^2 -1) \frac{1}{6} (A^A_{0\frac{1}{2}})^2
+\sqrt{\frac{2}{9}}(f^A_1(1) -1) 
A^A_{0\frac{1}{2}}\cdot A^A_{2\frac{3}{2}}
\nonumber \\
|H^A_{\frac{1}{2}1}|^2 \rightarrow |H^A_{\frac{1}{2}1}|^2 +
((f^A_1(1))^2 -1) \frac{1}{3} (A^A_{0\frac{1}{2}})^2
-\sqrt{\frac{2}{9}}(f^A_1(1) -1)
A^A_{0\frac{1}{2}}\cdot A^A_{2\frac{3}{2}}
\label{long}
\Ea     
When summing the two contributions (\ref{long}) in the rate formula the
$A^A_{0 \frac{1}{2}}\cdot A^A_{2 \frac{3}{2} }$
interference contributions cancel out as is apparent in (\ref{gen4}). As explained before we shall use the
leading order results (\ref{ls}) for the second and third term in (\ref{long})
since the factors $((f^A_1(1))^2 -1)$ and  $(f^A_1(1) -1)$ multiplying them
are already of ${\cal O}(1/m_Q^2)$.

%
%

Our numerical evaluation of (\ref{gen4}) is based on the standard form factor
(\ref{QCD}) with $\rho_B^2=0.75$ using again $V_{bc}=0.038$. 
All parameters have been specified before. For the quasi-elastic rate we find
$\Gamma^{{\rm excl}}= 5.52 \times 10^{10} s^{-1}$. The ${\cal O}(1/m_Q)$
and ${\cal O}(1/m_Q^2)$ corrections amount to $+5.2\%$ and $-6.6\%$.
The renormalization of the heavy quark current 
decreases the exclusive rate by $8.8\%$. 
In Fig.2 we show a plot of the $\om$-spectrum of the quasi-elastic rate
where we separately show the transverse ($\lambda_W=\pm1$)
and longitudinal contributions ($\lambda_W=0$) including  ${\cal O}(1/m_Q^2)$
and ${\cal O}(\alpha_s)$ corrections calculated according to (\ref{long}). The longitudinal rate dominates
the spectrum except for a small region close to zero recoil. For the integrated
rates we find $\Gamma_L^{{\rm excl}}$/$\Gamma_T^{{\rm excl}}= 1.89$.


\section{Inclusive Semileptonic Rate 
$\lowercase{\Lambda_b \rightarrow {\uppercase{X}}_c + l^- + 
\overline{\nu}_l}$}

To the leading order in the heavy mass expansion the inclusive rate is given by 
the free heavy quark decay rate which can be obtained from Eq.(\ref{gen})
using quark masses and setting the curly bracket equal to 1 . 
There are no ${\cal O}(1/m_Q)$ corrections to this result. 
Mass corrections come in at the order ${\cal O}(1/m_Q^2)$. In the case of 
$\Lambda_b$-decay, where the light diquark system has spin $0$,
the chromomagnetic contribution drops out and
the mass corrections 
are determined by the nonperturbative kinetic energy parameter 
$\mu_{\pi}^2$ alone. 

Including also the $\alpha_s$-correction in the free quark decay rate \cite{Nir}, 
one has ($x = (m_c/m_b)^2$) 
\Be
\Gamma^{\rm incl} = \Gamma_{0} \left ( 1 - \frac{2}{3} 
\frac{\alpha_s (m_b)}{\pi} g(x) \right )
\left ( 1 - \frac{\mu_{\pi}^2}{2 m_b^2} \right ) ,
\label{incl} 
\Ee
where $\Gamma_{0}$ is the lowest order (in $\alpha_s$) free quark decay rate
\Be
\Gamma_{0} = \frac{G_F^2 |V_{bc}|^2 m_b^5}{192 \pi^3} I_0(x) \; , 
\qquad\qquad I_0(x) =  (1-x^2)(1-8 x + x^2) -12 x^2 \ln x,
\label{born} 
\Ee
and the function $g(x)$ is determined by the ${\cal O}(\alpha_s)$  
radiative corrections including all mass corrections as calculated in \cite{Nir}:
\Ba
g(x) &=& h(x)/I_0(x)\; , \nonumber  \\
h(x) &=& -(1-x^2)\left ( \frac{25}{4} - \frac{239}{3} x + \frac{25}{4} x^2 \right )
+ x \ln(x)\left (20 + 90 x - \frac{4}{3} x^2 + \frac{17}{3} x^3 \right ) + 
x^2 \ln^2(x) 
(36 + x^2) \nonumber \\
&+&  (1-x^2)\left (\frac{17}{3} - \frac{64}{3} + \frac{17}{3}x^2 \right ) \ln 
(1-x) - 4 (1 + 30 x^2 + x^4)\ln (x) \ln (1-x) \nonumber \\
&-& (1 + 16 x^2 + x^4) [ 6 {\rm Li}_2 (x) - \pi^2 ] \nonumber \\
&-& 32 x^{3/2} (1+x) \left [ \pi^2 - 4 {\rm Li}_2 (\sqrt{x}) + 
4 {\rm Li}_2( -\sqrt{x}) - 2 \ln (x) \ln \left ( 
\frac{1-\sqrt{x}}{1+\sqrt{x}} \right ) \right ]
\label{g}
\Ea
The ${\cal O}(1/m_b^2)$ corrections appear 
in the third factor of Eq.(\ref{incl}).  
For the value of the kinetic energy parameter $\mu_{\pi}^2$ we take \cite{BB,PP}
\Be
\mu_{\pi}^2 = 0.5 \, \rm GeV^2 ,
\Ee
where we assume equality of the kinetic energy parameter in the meson and baryon case.

It is well known and evident from Eq.(\ref{born}) that the inclusive decay rate depends rather
strongly on the exact value of the $b$-quark mass $m_b$ which is fraught with some
uncertainties. We shall use the results of two recent theoretical analyses of the inclusive semileptonic
decay rate.  In \cite{PP} the value
of $m_b$ was determined from an analysis of $\Upsilon$ sum rules and $B$-meson semileptonic 
widths:
\Ba
m_b = 4.8 \,{\rm  GeV} , \;\; m_c = 1.325 \, {\rm  GeV}, 
\label{masses}
\Ea
where the charm quark mass was determined from the constraint
\Be
m_b - m_c - \mu_{\pi}^2 \left ( \frac{1}{2 m_c} -  
\frac{1}{2 m_b} \right ) = \overline{M}_B - \overline{M}_D .
\Ee
The $\overline{M}_{B,D}$ are the spin-averaged masses
$ \overline{M}_{B,D} = 1/4 ({M}_{B,D} + 3 {M}_{B^*,D^*})$ as before.

In \cite{HLM} the inclusive semileptonic $B$ decay rate was directly 
expressed in terms of the $\Upsilon (1S)$ meson mass instead of the $b$ quark 
mass.  
The authors of \cite{HLM} obtained: 
\Ba
\Gamma^{\rm incl - \Upsilon} = \frac{G_F^2 |V_{bc}|^2}{192 \pi^3}
\left( \frac{m_{\Upsilon}}{2} \right)^5 0.533 \left [ 1 - 0.096 \epsilon - 
0.029 \epsilon^2 - (0.28 \lambda_2 + 0.12 \lambda_1)/{\rm GeV^2} 
\right ]
\label{upsilon}
\Ea
where $\epsilon =1$ denotes the order of the expansion in $m_{\Upsilon}$. 
Mass corrections and radiative corrections are already taken into 
account. 
The parameters $\lambda_1$ and $\lambda_2$ in Eq.(\ref{upsilon}) 
are connected with 
the more familiar $\mu_{\pi}^2$ and $\mu_G^2$ parameters by 
$\mu_{\pi}^2 = -\lambda_1$ and $\mu_G^2 = 3 \lambda_2 = 0.36
 \; {\rm GeV}^2$.  For the $\Lambda_b$ baryon we set $\lambda_2 = 0$ and assume
equality of $\lambda_1$ in the meson and baryon case as before. 

For the semileptonic inclusive $b \rightarrow c$ decay rate of the  $\Lambda_b$ 
we finally obtain $\Gamma^{\rm incl} = 6.50\cdot 10^{10} s^{-1}$ using the mass
parameters from \cite{PP} and 
$\Gamma^{\rm incl - \Upsilon} = 6.23\cdot 10^{10} s^{-1}$ 
using the evaluation of
 \cite{HLM}. Again we have set $V_{bc}=0.038$. 
We mention that these two inclusive rate values
include the ${\cal O}(\alpha_s)$ radiative corrections which lower the inclusive rates by
about  $11\%$.  This value is not very far away from the  
 $8.8\%$ by which the exclusive
rate gets lowered by the same radiative corrections. 

\section{The Exclusive/Inclusive \lowercase{$\Lambda_b \rightarrow 
\Lambda_c$} Ratio}

In this section we determine the exclusive/inclusive ratio
$R_E=\Gamma^{\rm excl}/\Gamma^{\rm incl}$ in semileptonic  $\LL$ decays based on our 
estimates for the inclusive rates derived in Sec.IV and on  various 
phenomenological models 
for the baryonic Isgur-Wise function $F(\omega)$ entering in the exclusive differential 
rate Eq.(\ref{gen4}). Of all the phenomenological models we shall mostly focus
our attention on the sum rule calculation of  \cite{GY}.


We begin our discussion with the determination of the leading order
$\LL$ Isgur-Wise function
by the QCD sum rule method given in \cite{GY}.
The shape of the  Isgur-Wise function in \cite{GY} can be very well reproduced by an exponential
representation of the form
\Be
F(\om) =  \frac{2}{\om +1}\exp \left (-(2 \rho_B^2-1)\frac{\om -1}{\om +1}\right) , 
\label{QCD}
\Ee
which has the correct zero recoil normalization $F(1)=1$ and a slope parameter given
by $\rho_B^2$. The convexity parameter in this representation (proportional to $(\om -1)^2)$)  
is given by $c_B = 1/8(-1+4 \rho_B^2 + 4 \rho_B^4)$ and is positive for
$\rho_B^2 \ge 0.207$ as in most model calculations. 
We refer to this 
representation of the Isgur-Wise function as the standard form.
For the $\rho_B^2$ parameter the authors of  \cite{GY} find $\rho_B^2 = 0.85$
and $\rho_B^2 = 0.65$ using diagonal and nondiagonal sum rules, respectively.
 As an average of these two values one obtains 
\Be
\rho_B^2 = 0.75 .
\label{minQCD}
\Ee
Using the average value of $\rho_B^2$, $V_{bc}=0.038$, the standard representation of the Isgur-Wise
function (\ref{QCD}) and the rate formula (\ref{gen4}) from Sec.III one obtains
the exclusive rate $\Gamma^{\rm excl} = 5.52\cdot 10^{10} s^{-1}$. 
From the inclusive
rate calculated using the mass parameters given in \cite{PP}
$\Gamma^{\rm incl} = 6.50\cdot 10^{10} s^{-1}$ one
finds $R_E=0.85$ for the exclusive/inclusive ratio. Note that the 
$V_{bc}$ dependence drops out in this ratio. The values of the slope parameter
$\rho_B^2$ and the exclusive/inclusive ratio $R_E$ of the model of  \cite{GY} as well
as those of other phenomenological models have been collected together in Table I.
Table I uses the  larger value of the two inclusive reference rates 
discussed in Sec.IV based on the mass parameters of \cite{PP}. If one instead uses the
inclusive rate of  \cite{HLM} all $R_E$-values in Table I have to be increased by 4.3\%. 
Radiative corrections do not affect the exclusive/inclusive ratios listed in Table I very much
since they lower both the exclusive and inclusive rates (see Secs. III and IV). If they were
left out the exclusive/inclusive ratio $R_E$ would be reduced by $\approx 2\%$.


It is clear from Table I that the form factor calculated in the quark confinement model \cite{EIKL}
is too flat to satisfy the bound $\Gamma^{\rm excl} \le \Gamma^{\rm incl}$. All other models
in Table I satisfy this upper bound. Translating the upper bound $R_E=1$ into a lower bound
on the slope parameter $\rho_B^2$ one obtains    
\Be
(\rho_B^2)_{\rm min} = 0.36 
\label{min}
\Ee
using the standard form factor function (\ref{QCD}) and the inclusive rate calculated from the mass
parameters in \cite{PP}. 



An upper bound on the slope parameter $\rho_B$ can be obtained using the spectator model 
bound discussed in Sec.II, which reads 
\Be
\rho_B^2 \le 2 \rho_M^2 -\frac{1}{2} .
\label{bm}
\Ee
The mesonic slope parameter $\rho_M^2$ can be 
extracted from the exclusive semileptonic $B$ decays \cite{Drell}. 
The values are
\Ba
(\rho_M^2)_1 &=& 0.66 \pm 0.19 \qquad\qquad\qquad {\rm from} 
\qquad \overline{B} \rightarrow D l^- \overline{\nu}
\nonumber \\ 
(\rho_M^2)_2 &=& 0.71 \pm 0.11 \qquad\qquad\qquad {\rm from} 
\qquad \overline{B} \rightarrow D^* l^- \overline{\nu}
\label{exper}
\Ea
with the world average values for $V_{cb}$ being 
$|V_{cb}| = 0.0394 \pm 0.0050$ and 
$|V_{cb}| = 0.0387 \pm 0.0031$, respectively.
The weighted average of the two mesonic slope parameters are then
$\rho_M^2=0.70 \pm 0.10$. This translates into an upper bound for
the baryonic slope parameter $\rho_B^2$ according to
the spectator quark model bound (\ref{bm}). One has
\Be
(\rho_B^2)_{\rm max} =0.89 \pm 0.19 \, .
\label{max}
\Ee
We mention that very likely the error on this bound will be considerable reduced in the near
future with the new data expected from the bottom factories at SLAC and KEK.
Combining both limits, (\ref{min}) and (\ref{max}), we obtain a 
prediction for the allowed values of the baryon slope parameter given by
\Ba
0.36 < \rho_B^2 < 0.89 \pm 0.19. 
\label{bound}
\Ea
\begin{table}[t]
\caption{Predictions of the slope parameter $\rho^2_B$ and the ratio
$R_E=\Gamma^{\rm excl}/\Gamma^{\rm incl}$ calculated in different models.}
\begin{tabular}{cccc}
\\
MODEL & ISGUR-WISE FUNCTION &  $\rho^2_B$   &  $\Gamma^{\rm excl}/\Gamma^{\rm incl}$ \\
\\
\hline
\\
quark confinement $\rm model^{\cite{EIKL}}$ & $\frac{\ln(\om + \sqrt{\om^2 -1})}{\sqrt{\om^2-1}}$ & 0.33 & 1.02 \\
\\
QCD sum $\rm rules^{\cite{DHHL}}$ & $\sqrt{\frac{2}{\om +1}}\exp\left (-0.8 \frac{\om -1}{\om +1}\right)$ &
0.65 & 0.89 \\
\\
QCD sum $\rm rules^{\cite{GY}}$($\rho^2=0.75$)  & $\frac{2}{\om +1}\exp \left (-(2 \rho^2-1)\frac{\om -1}{\om +1}\right)$ &
0.75 & 0.85 \\
\\
Simple quark $\rm model^{\cite{HSM}}$ & $\left (\frac{2}{\om +1 }\right )^{(1.32 + \frac{0.7}{\om})}$ &   1.01 & 0.78 \\
\\
Relativistic three quark $\rm model^{\cite{kroll}}$ & $\left (\frac{2}{\om +1 }\right )^{(1.7 + \frac{1}{\om})}$ &   1.35 & 0.68 \\
\\
IMF $\rm model^{\cite{KKKK}}$ & $\frac{1}{\om}
\exp \left (-(0.7)^2 \,\frac{\om -1}{2 \om} \right ) \cdot \frac{
\int_{-0.7 \sqrt{\frac{\om +1}{2 \om}}}^{\infty}
dy e^{-y^2} (y + 0.7 \sqrt{\frac{\om +1}{2 \om}}) }{
\int_{-0.7}^{\infty}
dy e^{-y^2} (y + 0.7)} $ &  1.44 & 0.66 \\
\\
Skyrme model in the large $N_c$ $\rm limit^{\cite{JMW}}$  & $0.99 \exp ( -1.3 (\om -1))$ & 1.30 & 0.63 \\
\\
MIT bag $\rm model^{\cite{SZ}}$ & $\left (\frac{2}{\om +1 }\right )^{(3.5 + \frac{1.2}{\om})}$ &   2.35 & 0.45 \\
\\
\end{tabular}
\label{t:tab1}
\end{table}
According to these upper and lower bounds the first model (as remarked on before)
and the last four models in Table I have
to be excluded
since they possess form factors which are too flat or too steep, respectively.
The two QCD sum rule calculations \cite{DHHL,GY} as well as the simple quark model evaluation
\cite{HSM} feature slope parameters that satisfy the bounds (\ref{bound}).
We consider the two QCD sum rule calculations to be the most reliable of the three
model calculations since they are the least model dependent. Our final prediction for the range
of values of the exclusive/inclusive ratio will be based on the two slope parameter values
$\rho_B^2=0.85$ and $0.65$, resulting from the analysis of the diagonal and nondiagonal
\cite{GY} sum rules in \cite{GY}, respectively. This range also includes the sum rule result
of \cite{DHHL}. In determining our prediction for the range of $R_E$
we shall also allow for the smaller inclusive rate calculated by the method of \cite{HLM}.
Thus our final prediction for the range of the exclusive/inclusive ratio in semileptonic
$\LL$ decays is $R_E= 0.81 \div 0.92$. 
This range is consistent with 
the range of values from
the semiquantitave
analysis performed in Sec.II. Our conclusion is that the exclusive/inclusive  
ratio of semileptonic $\Lambda_b$-decays
is considerable higher than
in the corresponding bottom meson case. 

\section{Missing final states}

Besides the quasi-elastic $\LL$ contribution discussed before there are
also  $\Lambda_c^{\ast \ast}$ resonant states and multiparticle final states
contributing to the fully inclusive
semileptonic $\Lambda_b$ rate. Of course, if the quasi-elastic
contribution dominates the total inclusive rate much more than by the 66\%
in the heavy meson case, there would not be much room left for the
resonant and multiparticle final states. In the main body of this paper we have
collected together theoretical evidence that the latter situation is very likely the case.
One could turn this statement  around in the 
following sense: if one would have theoretical reasons to believe that resonant and
multiparticle final states are suppressed in inclusive semileptonic
$\Lambda_b \rightarrow X_c$ transitions then
the quasi-elastic exclusive $\LL$ contribution must dominate. As we shall see there are
theoretical reasons to believe in such a suppression in as much as some of the transitions
to orbitally excited $\Lambda_c^{\ast \ast}$  charm baryon states involve spin-orbit
coupling transitions which are believed to be suppressed.

The purpose of this section is to classify those final states in semileptonic
$\Lambda_b \rightarrow X_c$ transitions that form the complement of the 
quasi-elastic $\LL$ transition. We divide these into class $A$ contributions
$\LL X l \nu$, where the charm quark of the
decay ends up in a charm $\Lambda_c$ directly or indirectly, and class $B$
contributions, where the charm quark goes into a meson or a 
charm-strangeness baryon $\Lambda_b \rightarrow X_c({non}\,\Lambda_c) l\nu$.
Accordingly we define the two ratios
\Be
R_A=\frac{\Gamma (\LL X l  \nu )}{\Gamma (\Lambda_b \rightarrow X_c l \nu)} .
\Ee
and
\Be
R_B=\frac{\Gamma (\Lambda_b \rightarrow X_c({non}\, \Lambda_c) l \nu )}
{\Gamma (\Lambda_b \rightarrow  X_c l \nu)} . 
\Ee
Together with the exclusive/inclusive ratio defined before
\Be
R_E=\frac{\Gamma (\LL l  \nu )}{\Gamma (\Lambda_b \rightarrow X_c l \nu)} ,
\Ee
the three ratios must add up to one, {\it i.e.}
\Be
R_E + R_A +R_B = 1.
\label{con}
\Ee
Note that all three ratios are positive definite which makes the constraint (\ref{con})
potentially quite powerful if $R_E$ is close to one as is indicated by our analysis of the
quasielastic rate in the previous sections. As concerns the sizes of $R_A$ and
$R_B$ one cannot even hope to provide semiquantitative answers at present. 
It is nevertheless useful to enumerate the final states belonging to the
class $A$ and class $B$ transitions which we shall do in the following. 
\\

\underline{Class $A$ final states} \\

Potentially prominent among the class $A$ final states are the transitions into the seven 
excited $P$-wave $\Lambda_c^{\ast  \ast}$-states. Taking the bottom meson case
for comparison theoretical estimates show that the corresponding transitions into
excited mesonic $P$-wave states
make up approximately 10\% of semileptonic B-decays \cite{vesely}.  The
$\Lambda_c^{\ast \ast}$-states eventually decay down to the $\Lambda_c$
ground state via (multiple) pion emission or, with a much smaller 
branching fraction, via  photon emission. There are altogether seven such $P$-wave
states which are grouped into the three
HQS doublets 
$\{ \Lambda_{cK1} \}$, $\{ \Lambda_{ck1} \}$, $\{ \Lambda_{ck2} \}$, and the 
singlet $\{ \Lambda_{ck0} \}$. We use the terminology of \cite{KKP} such
that the excited $K$- and $k$-states are symmetric and antisymmetric under
the exchange of the momenta of the light quarks. The five symmetric states
$\{ \Lambda_{ck0} \}$, $\{ \Lambda_{ck1} \}$, 
and $\{ \Lambda_{ck02} \}$ are made from a heavy quark and a light spin-one diquark.
$\Lambda_b$ transitions into these five states involve spin-zero to spin-one
light-side transitions which can be expected to be strongly suppressed since they
involve spin-orbit interactions.
In the spectator quark model, where one neglects spin-orbit interactions, transitions into
these five states are forbidden \cite{hussain}. It would be interesting to experimentally confirm this
suppression. One thus remains with the transitions into the 
HQS doublet $\{ \Lambda_{cK1} \}$ whose spin $1/2^-$ and spin $3/2^-$ members are very likely 
the recently discovered $\Lambda_c (2593)$ and  $\Lambda_c (2625)$ states \cite{PDG}. 
$\Lambda_b$ branching ratios into these states are not yet available.
There could also be transitions into higher
orbital $\Lambda_c^{\ast \ast}$-states. These transitions are, however,
expected to be suppressed because of angular momentum suppression factors. Besides,
transitions into symmetric higher orbital $\Lambda_c^{\ast \ast}$-states would again
be suppressed due to spin-orbit coupling suppression. The suppression of transitions
into the symmetric orbitally excited $\Lambda_c^{\ast \ast}$-states could be the source
of the possible depletion of class $A$ final states. For example, using spin counting only
$1/3$ of the existing $P$-wave excitations can be reached in semileptonic
$\Lambda_b$ transitions if the spin-orbit coupling suppression is active. 

\begin{figure}[h]
  \centerline{\epsfig{file=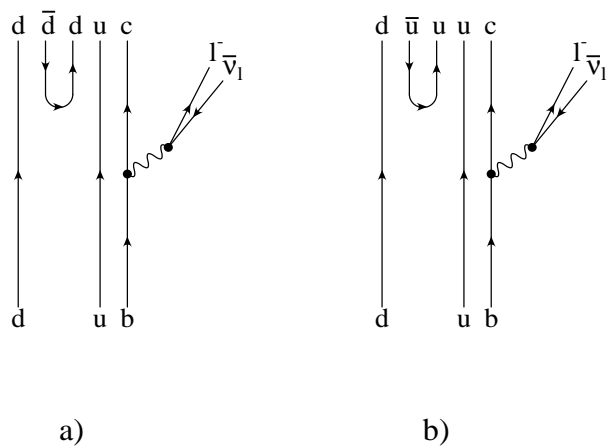,height=6cm,width=8.5cm,silent=}}
 \caption{Class $A$ final states.}
 \label{f:fig3}
\end{figure}

Another source of  class $A$ final states is accessible due to the creation
of one or more additional 
$(d \overline{d})$- or $(u \overline{u})$-quark pairs in the basic transition. 
The relevant 
transitions for $(d \overline{d})$ creation are (see Fig.3a)
\Ba
\Lambda_b^0 &\rightarrow& \Lambda_c^+ (\Sigma_c^+) + X_M^0 + l^- + 
\overline{\nu}_l , 
\label{A1}
\Ea
or, when exchanging the $d \leftrightarrow u$ lines originating from the 
$\Lambda_b$, one has
\Ba
\Lambda_b^0 &\rightarrow& \Sigma_c^0 + X_M^+ + l^- + 
\overline{\nu}_l .
\label{A2}
\Ea
For $(u \overline{u})$ creation shown in Fig.3b one has
\Ba
\Lambda_b^0 &\rightarrow& \Sigma_c^{++} + X_M^- + l^- + 
\overline{\nu}_l . 
\label{A3}
\Ea
The  exchange of the $d,u$ lines originating from the $\Lambda_b$ brings 
one back to (\ref{A1}).
$X_M$ stands for a charmless mesonic inclusive state. 
Excited charm baryon states such as
$\Lambda_c^{**}$ and $\Sigma_c^{**}$  are not explicitly included in the listing 
(\ref{A1} - \ref{A3}), but are implied. The $\Sigma_c^0$, $\Sigma_c^+$, and
$\Sigma_c^{++}$ appearing in
(\ref{A1}-\ref{A3}) cascade down to the $\Lambda_c^+$ state 
via pion emission making these processes class $A$ final states.
\\

\underline{Class $B$ final states} \\

There are two sources for class $B$ final states. 
First there is $(s \overline{s})$ quark pair creation where the strange
quark ends up  in a charm-strangeness baryon (Fig.4a) which decays weakly
into noncharm states and therefore does not contribute to the class 
$A$ final states
\footnote{The weak decay $\Xi_c \rightarrow \Lambda_c + \pi$,
though interesting, occurs only at the per mill level \cite{voloshin}.}. 
Second, the charm quark of the decay may end up in a charm meson accompanied by
$(u \overline{u})$-, $(d \overline{d})$, and $(s \overline{s})$-quark pair 
creation as shown in Figs.4b-d.  Let us list a few examples of such
transitions. From $(s \overline{s})$ pair creation one has (Fig.4a)
\Ba
\Lambda_b^+ &\rightarrow& \Xi_c^+ + X_{M_s}^0 + l^- + \overline{\nu}_l ,
\Ea
or, when exchanging the $d \leftrightarrow u$ lines, one has
\Ba
\Lambda_b^+ &\rightarrow& \Xi_c^0 + X_{M_s}^+ + l^- + \overline{\nu}_l ,
\label{B1}
\Ea
$X_{M_s}$ now stands for a strangeness meson state. Then there are the
transitions where the charm quark goes into a charm meson. These are

\begin{figure}[h]
  \centerline{\epsfig{file=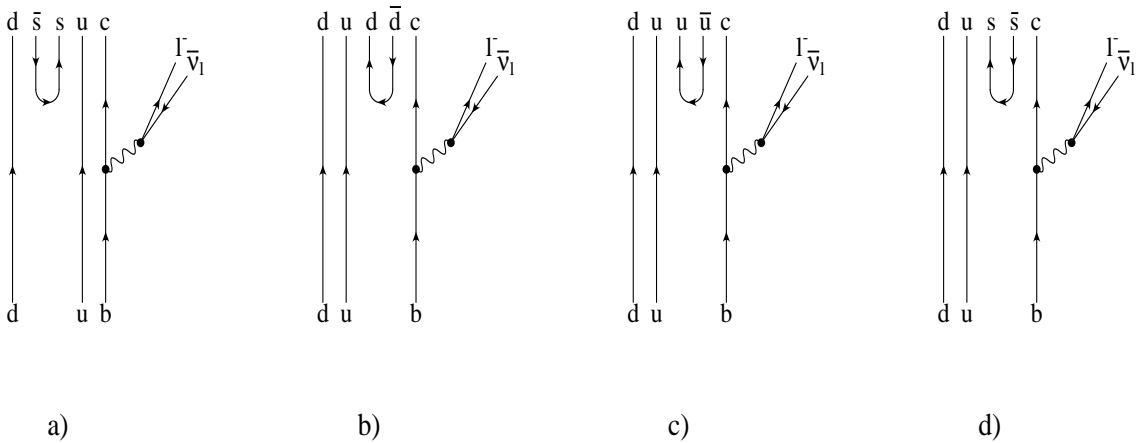,height=6cm,width=15.5cm,silent=}}
 \caption{Class $B$ final states.}
 \label{f:fig4}
\end{figure}

\Be
\Lambda_b^+ \rightarrow D^+ + X_B^0 + l^- + \overline{\nu}_l , 
\label{B2}
\Ee

\Be
\Lambda_b^+ \rightarrow D^0 + X_B^+ + l^- + \overline{\nu}_l , 
\label{B3}
\Ee

\Be
\Lambda_b^+ \rightarrow D_s^+ + X_{B_s}^0 + l^- + \overline{\nu}_l , 
\label{B4}
\Ee
$X_B$ stands for a light baryon state and $X_{B_s}$ for
a strangeness baryon state. Excitations of the charm meson and charm 
baryon states are again implied.
We do not discuss $(c \overline{c})$ pair creation. 
The corresponding final states are barely
accessible in semileptonic $\Lambda_b$ decays for 
kinematical reasons and will have a spectacular signature anyhow.

\section{Summary and Conclusion}
          
We have brought together various pieces of theoretical evidence
that the exclusive/inclusive ratio $R_E$ in semileptonic
$\Lambda_b$-decays is larger than in semileptonic $B$ decays,
where the exclusive/inclusive ratio amounts to 66\%. We predict that the exclusive
quasielastic semileptonic
$\Lambda_b$-decays make up
between 81\% and 92\% of the total inclusive semileptonic $\Lambda_b$ rate.  At present there
is no experimental information on either 
the exclusive or the inclusive branching ratio  in semileptonic $\Lambda_b$ decays.
The problem is that present and
planned experiments do not have access to reliable $\Lambda_b$ tags which are
necessary for a 
measurement of their branching fractions. Ideally one would run a
$e^+ e^-$-machine right above $\Lambda_b \overline{\Lambda}_b$ threshold
which would solve the tagging problem. However, such experiments are not
planned in the foreseeable future. The above assertion about the dominance
of  the quasi-elastic mode in semileptonic $\Lambda_b$ decays may take a long
time to verify experimentally. It may nevertheless be used as a working hypothesis
in the experimental analysis of semileptonic $\Lambda_b$ decays in particular if
further theoretical progress in the theoretical description of 
semileptonic $\Lambda_b$ decays confirms the estimates made in this paper.
 


\acknowledgements

This investigation was prompted by two questions of our experimental colleagues
P. Roudeau and G. Sciolla who asked us about theoretical expectations for the
size of the the $\Lambda_c +X$ (class A) and $({non}\Lambda_c) + X_c$ (class B)
contributions in inclusive
semileptonic $\Lambda_b$ decays. We did our best to try and provide
answers to these questions in our paper, at least partially. We would like to 
thank N. Uraltsev for fruitful discussions. We also thank O. Yakovlev who participated
in the early stages of this calculation. 
The work of B.M. was partially supported by the Ministry of
Science and Technology of the Republic of Croatia under the contract Nr.
00980102.

\end{document}